\documentclass[twocolumn]{aastex631}
\usepackage[normalem]{ulem}
\newcommand{\src}{Her~X-1}

\begin{document}

\title{Flux-resolved Spectropolarimetric Evolution of the X-Ray Pulsar Hercules X-1 Using {\it{IXPE}}}

\author{Akash Garg}\thanks{E-mail: akash.garg@iucaa.in (AG)}
\affiliation{Inter-University Center for Astronomy and Astrophysics, Ganeshkhind, Pune 411007, India}

\author{Divya Rawat}
\affiliation{Inter-University Center for Astronomy and Astrophysics, Ganeshkhind, Pune 411007, India}

\author{Yash Bhargava}
\affiliation{Department of Astronomy and Astrophysics, Tata Institute of Fundamental Research, 1, Homi Bhabha Road, Colaba, Mumbai 400005, India}

\author{Mariano M\'endez}
\affiliation{Kapteyn Astronomical Institute, University of Groningen, PO BOX 800, Groningen NL-9700 AV, the Netherlands}
\author{Sudip Bhattacharyya}
\affiliation{Department of Astronomy and Astrophysics,  Tata Institute of Fundamental Research, 1, Homi Bhabha Road, Colaba, Mumbai 400005, India}

\begin{abstract}
We conduct a spectro-polarimetric study of the accreting X-ray pulsar Hercules X-1 using observations with the \textit{Imaging X-ray Polarimetry Explorer} (\textit{IXPE}). \textit{IXPE} monitored the source in three different Epochs, sampling two Main-on and one Short-on state of the well-known super-orbital period of the source. We find that the $2-7$~keV polarization fraction increases significantly from $\sim 7-9 \%$ in the Main-on state to $\sim 15-19 \%$ in the Short-on state, while the polarization angle remains more or less constant or changes slightly, $\sim 47-59$ degrees, in all three Epochs. The polarization degree and polarization angle are consistent with being energy-independent for all three Epochs. We propose that in the Short-on state, when the neutron star is partially blocked by the disk warp, the increase in the polarization fraction can be explained as a result of the preferential obstruction of one of the magnetic poles of the neutron star.

\end{abstract}

\section{Introduction} \label{sec:intro}
Hercules X-1 (\src) is an accreting X-ray binary pulsar with a low mass companion, HZ Her, discovered in the 1970s \citep{sh72,ta72}. The light curve of \src\ shows pulsations at the neutron-star (NS) spin period of 1.24~s \citep{Staubert2013A&A...550A.110S} and clear dips due to occultation by the companion at 1.7~d intervals \citep{Staubert2013A&A...550A.110S,2014ApJ...793...79L}, indicating that the source has a high inclination ($>$80$^\circ$, \citealt{de81}). Additionally, the source is known to show a super-orbital period of 34.85 days \citep{giacconi1973ApJ...184..227G}, which is assumed to occur due to a warp in the accretion disk around the NS \citep{pe91,2002MNRAS.334..847L}. The super-orbital period is characterized by two high flux states, `Main-on' and `Short-on', with two `Off' states in between \citep{Staubert2013A&A...550A.110S,2020ApJ...902..146L}. 

The broadband spectrum of \src\ is known to arise from multiple regions around the neutron star \citep{2007ApJ...654..435B,2015MNRAS.453.4222A,kosec2022ApJ...936..185K}. The hard X-ray pulsating emission ($>$ 2~keV) originates from the accretion column, while the soft X-ray emission ($<$ 1~keV) comes from the reprocessed flux of the accretion column radiation. The column density of the interstellar matter toward \src\ is low \citep[5$\times10^{19}$~cm$^{-2}$;][]{1998A&A...329L..41D}, and the spectrum of the source exhibits an evolving cyclotron resonant scattering feature \citep[$35-42$~keV;][]{gruber2001ApJ...562..499G, furst2013ApJ...779...69F, ji2019MNRAS.484.3797J, bala2020MNRAS.497.1029B}, indicative of the high NS magnetic field.  \src\ exhibits a highly asymmetric pulse profile that is dependent on the superorbital period \citep{leahy2004ApJ...613..517L}. The evolution of the pulse phase in the Main-on and the Short-on state suggests that the emission from the closest magnetic pole is a pencil beam whereas that of the farthest pole is a fan beam
\citep{scott2000ApJ...539..392S, leahy2004ApJ...613..517L}. \\

Although the X-ray spectral and temporal properties of \src\ have been extensively studied to investigate the configuration and emission regions, there remains a degeneracy of the models. Through polarimetric observations, it is possible to delve into the geometry of the accretion column and magnetic field and examine the changes in polarization during the super-orbital phase. Recent work by \citet{doroshenko2022NatAs...6.1433D} has explored the soft X-ray polarization properties of \src\ using the {\it{Imaging X-ray Polarimetry Explorer}}, {\it{IXPE}}, during the Main-on state (phase range $0.0-0.2$ of the super-orbital period), revealing an energy-independent polarization in the $2-7$~keV range that strongly depends on the pulse phase. They report a significant X-ray polarization degree (PD) of 8.6 $\pm$ 0.5\% at a polarization angle (PA) of $62^\circ \pm 2$, which
was much lower than the theoretically expected value \citep{ca21}. They further speculated three possible reasons for this low PD value: (1) radiative transfer in the magnetized plasma within the emission region; (2) propagation of the initially polarized X-rays through the NS magnetosphere; and (3) the effect of the combination of the emission from the two magnetic poles.
\\

In this letter, we present a comparison of the polarization properties of \src\ in the Main-on and Short-on states of the super-orbital phase as seen by {\it{IXPE}} at three different Epochs. The goal of this study is to explore how the accretion disk warping affects polarization.
The methods and data analysis procedures for the observations are outlined in Section~\ref{sec_obs}, and the results and ensuing discussion are reported in Sections \ref{Sec_res} and \ref{sec_concl}, respectively. 

\section{Observation and Data Analysis}\label{sec_obs}

{\it{IXPE}} \citep{we22} is NASA's dedicated X-ray polarimetry mission, launched on 2021 December 9 from the Kennedy Space Center. It has onboard three identical units of Gas Pixel detectors, DU1, DU2 and DU3 \citep{ba21}. These units independently record the spatial, time, and energy-resolved polarimetric information in the $2-8$~keV band \citep{di22}. \textit{IXPE} observed the X-ray pulsar \src\ at three different Epochs, 2022 February 17-24, 2023 January 18-21, and 2023 February 3-8. The observation IDs for all Epochs are given in Table~\ref{PA_PD_obs_log}. If not stated otherwise, all reported errors indicate the $1-\sigma$ confidence (68\%) range for the associated parameter.\\ 

\begin{figure}
% \hspace{-1.0cm}
\centering\includegraphics[scale=0.22,angle=0]{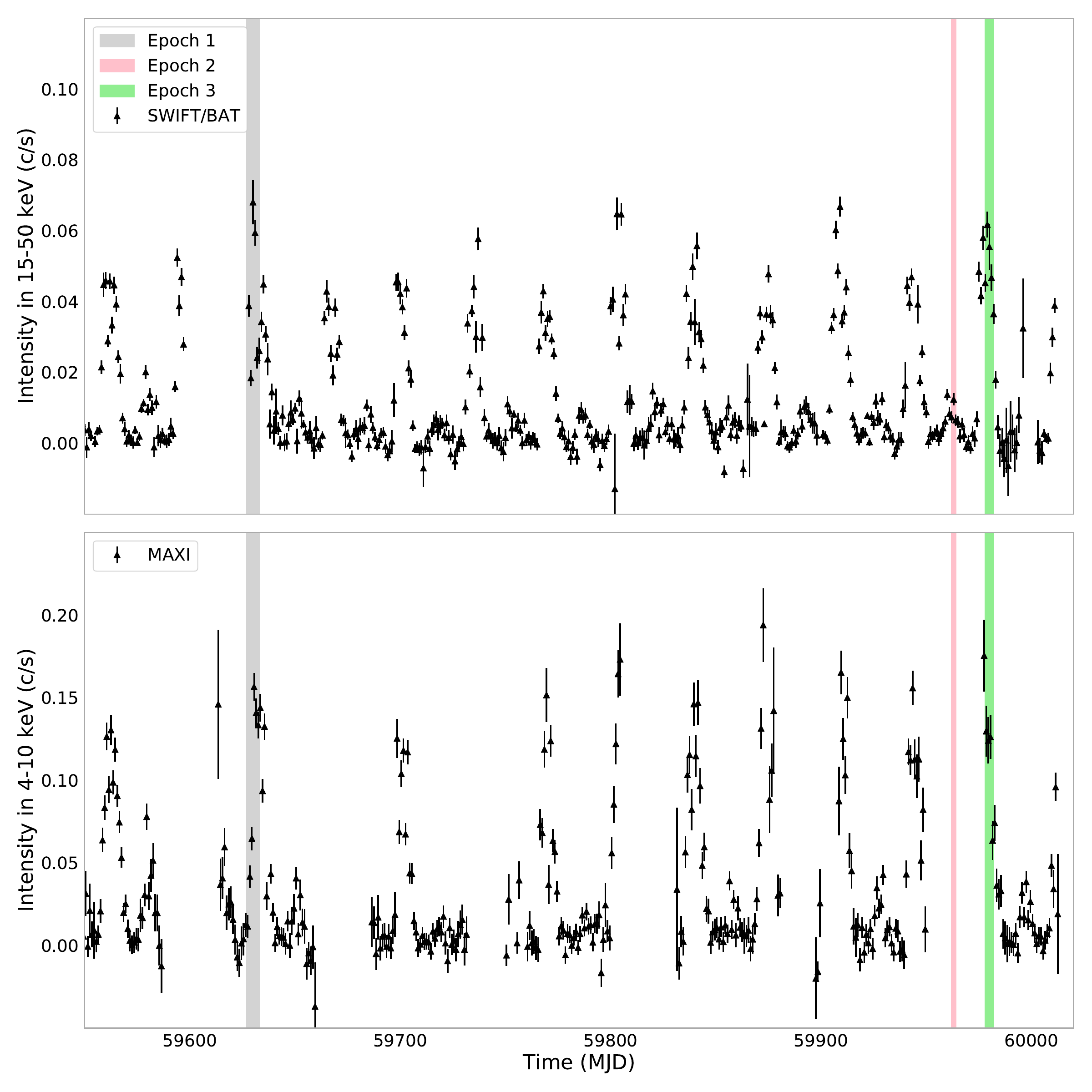}
\caption{{\it{Swift/BAT}} (upper panel) and  {\it{MAXI}} (lower panel) light curve of \src\ in the $15-50$~keV, and $4-10$~keV band, respectively. The shaded area represents the duration of the {\it{IXPE}}  observations of the source.}
\label{lightcurve}
\end{figure}

The upper and lower panels of Figure~\ref{lightcurve} show the \textit{Swift/BAT} and \textit{MAXI} lightcurves of \src\ in the $15-50$~keV and $4-10$~keV energy bands, respectively. Both lightcurves exhibit the well-known $\sim$35 day super-orbital period of the source. In the first and third {\it{IXPE}} Epochs, the source was in the Main-on state (grey and green shaded regions in Figure~\ref{lightcurve}, respectively), which marks the start of the super-orbital cycle. In the second {\it{IXPE}} Epoch, \src\ was detected in the Short-on state (pink shaded region in Figure~\ref{lightcurve}). Throughout each Epoch, there are brief intervals of eclipsing by the donor star and pre-eclipsing dips caused by obscuration due to the accretion stream from the donor star \citep{2012MNRAS.425....8I,doroshenko2022NatAs...6.1433D}. We discard all such dips to conduct the analysis and generate multiple segments for each Epoch, facilitating a time-dependent investigation of the polarization properties. Further, we extract a merged cleaned event file (to combine the  different segments to create a single event file per Epoch) using the {\sc{heasoft}} tool {\sc{xselect}} to determine the time-averaged polarization properties.

The polarimetric analysis is carried out using the software suite {\sc{ixpeobssim}} 29.2.0\footnote{\label{note2}\url{https://ixpeobssim.readthedocs.io/en/latest/overview.html}} \citep{ba22}, which is equipped with various tools to process {\it{IXPE}} level2 event files and produce scientific results. For each individual segment and for the merged event file in each Epoch, we procure the source and background cleaned event files for all DUs using the {\sc{xpselect}} tool. We choose a circular region (radius of 60'') and an annular region (inner and outer radii of 180'' and 240'') for the extraction of source and background events, respectively. Based on the formalism developed by \cite{2015APh....68...45K}, the tool {\sc{xpbin}} assists in binning the event files to generate the polarization cubes, count \textit{I} and Stokes \textit{Q} and \textit{U}, spectra using algorithms like {\sc{pcube}}, {\sc{pha1}}, {\sc{phaq}} and  {\sc{phau}}, respectively (For more details, also see \citealt{2022hxga.book....6M,2023arXiv230302745R}). Owing to uncertainties in the $7-8$~keV band \citep{doroshenko2022NatAs...6.1433D}, all the binned products are estimated in the $2-7$~keV energy range using the response matrices version v012 of {\sc{ixpeobssim}}.\\

Using the {\sc{heasoft}} tool {\sc{ftgrouppha}}, we have rebinned the \textit{I}, \textit{Q} and \textit{U} spectra. Specifically, we first rebin the \textit{I} spectrum uniformly by a factor of 2 for Epochs 1 and 3 and by a factor of 3 for Epoch 2, and then rebin the Q and U spectra using the binning of the \textit{I} spectrum as a template. The spectro-polarimetric fitting is conducted using the X-ray spectral fitting package {\sc{xspec version 12.13.0}} \citep{ar96}. \citet{doroshenko2022NatAs...6.1433D} remarked in their work that during Epoch 1, there is an additional vignetting due to the offset in the pointing direction of \textit{IXPE} telescopes and due to uncertainties in boom motion modeling which can influence effective area calibration and thereby the spectral analysis. They also noted that this would not affect the polarization measurements as the Stokes spectra \textit{U} and \textit{Q} are normalized by the \textit{I} flux which in turn make up for the systematic uncertainties coming from the effective area. No information on offset or other uncertainties is available for Epoch 2 and 3 data.

%%%%%%%%%%%%%%%%%%%%%%%%%%%%%%%%%%%%%%%%%%%
%%%%%%%%%%%%%%%%%%%%%%%%%%%%%%%%%%%%%%%%%%%%%%%%%%

%%%%%%%%%%%%%%%%%%%%
\begin{table}
\centering
 \caption{Log of the observations with the time-averaged PA and PD for \src\ with {\sc{pcube}} and {\sc{xspec}} model {\sc{constant*polconst*nthcomp}} for all the three Epochs.}
% \begin{center}
\scalebox{0.92}{%
\hspace{-1.0cm}
\begin{tabular}{ccccc}
\hline
Epoch & ObsID & parameter & {\sc{pcube}} &  {\sc{xspec}} \\
      &     &      & All DUs  &  \\
\hline
1 &01001899      & PA (degree) &   $ 59.2 \pm 1.5$ & $59.3 \pm 1.7$\\ 
         &     & PD (\%)       & $7.5 \pm 0.4$  & $8.5 \pm 0.5$\\     
\\~\\
2 &02003801      & PA (degree)   &  $46.6 \pm 3.4$ & $44.2 \pm 2.7$\\ 
         &     & PD  (\%)      & $17.3 \pm 2.0$   & $19.0 \pm 1.8$\\ 
\\~\\
3 & 02004001     & PA (degree)   & $49.8 \pm 1.9$  &  $50 \pm 1.8$\\ 
        &      & PD  (\%)      & $8.2 \pm 0.5 $   & $7.8 \pm 0.5$\\ 
      \hline
\end{tabular}}
\label{PA_PD_obs_log}
\end{table}
%%%%%%%%%%%%%%%%%%%%%%%%%%%%%%%%%%%%%%%%%%%
 \begin{figure}
% \hspace{-1.0cm}
\centering\includegraphics[width=\columnwidth,angle=0]{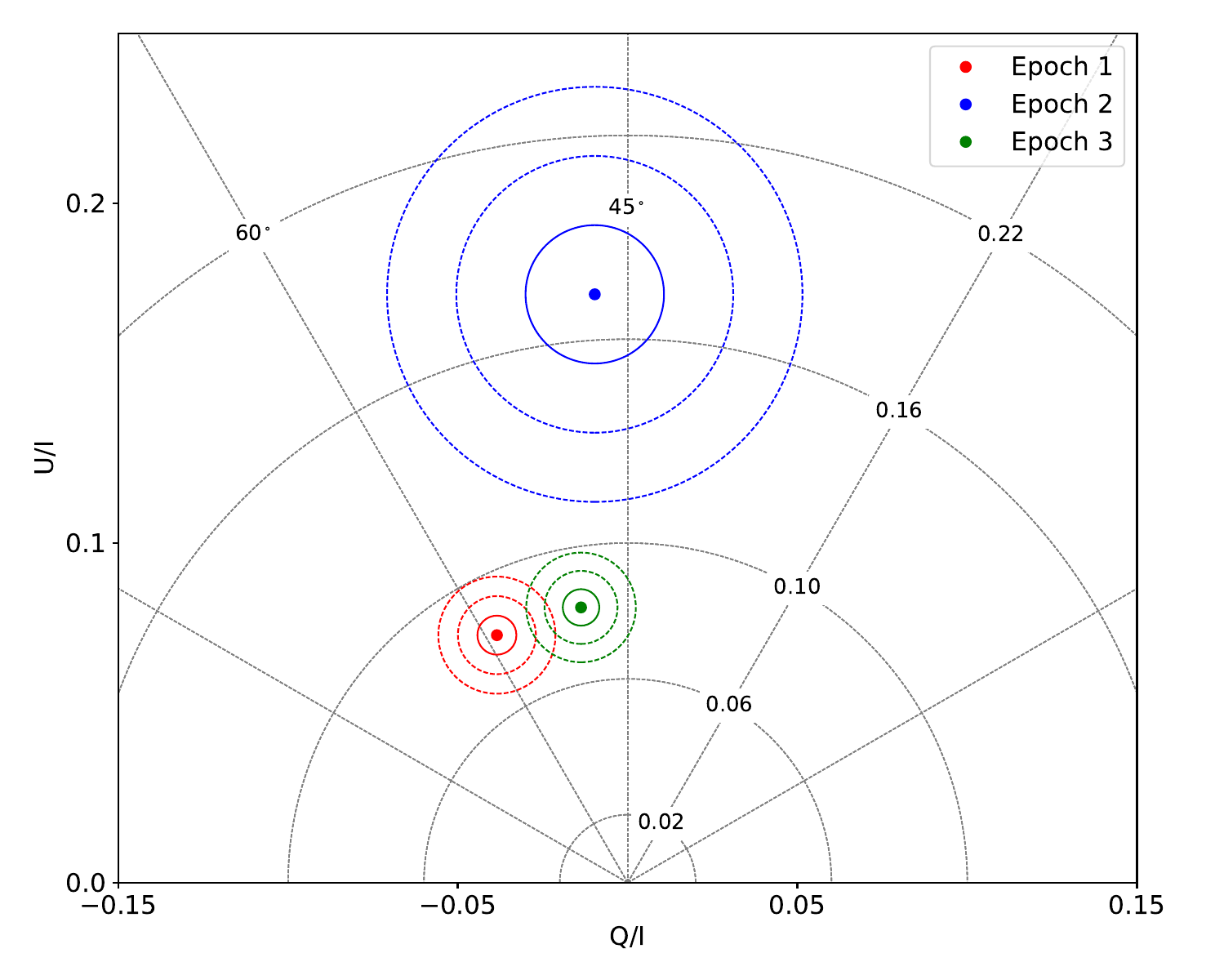}
\caption{The PD (azimuthal lines), PA (radial lines), and normalized Stokes parameters, $Q/I$ (X-axis) and $U/I$ (Y-axis), of \src\ for the three Epochs (please see legends) with {\it IXPE}. The measurements were obtained using the {\sc{pcube}} tool in the $2-7$~keV band. The contours are at 68.27, 95.45, and 99.73 \% confidence levels for both Stokes parameters.}
\label{4dparam}
\end{figure}
%%%%%%%%%%%%%%%%%%%%%%%%%%%%%%%%%%%%%%%%%%%%%%%%%%
%%%%%%%%%%%%%%%%%%%%%%%%%%%%%%%%%%%%%%%%%%%%%%
\begin{figure*}
\centering
\includegraphics[width=8.5cm,height=6cm,angle=0]{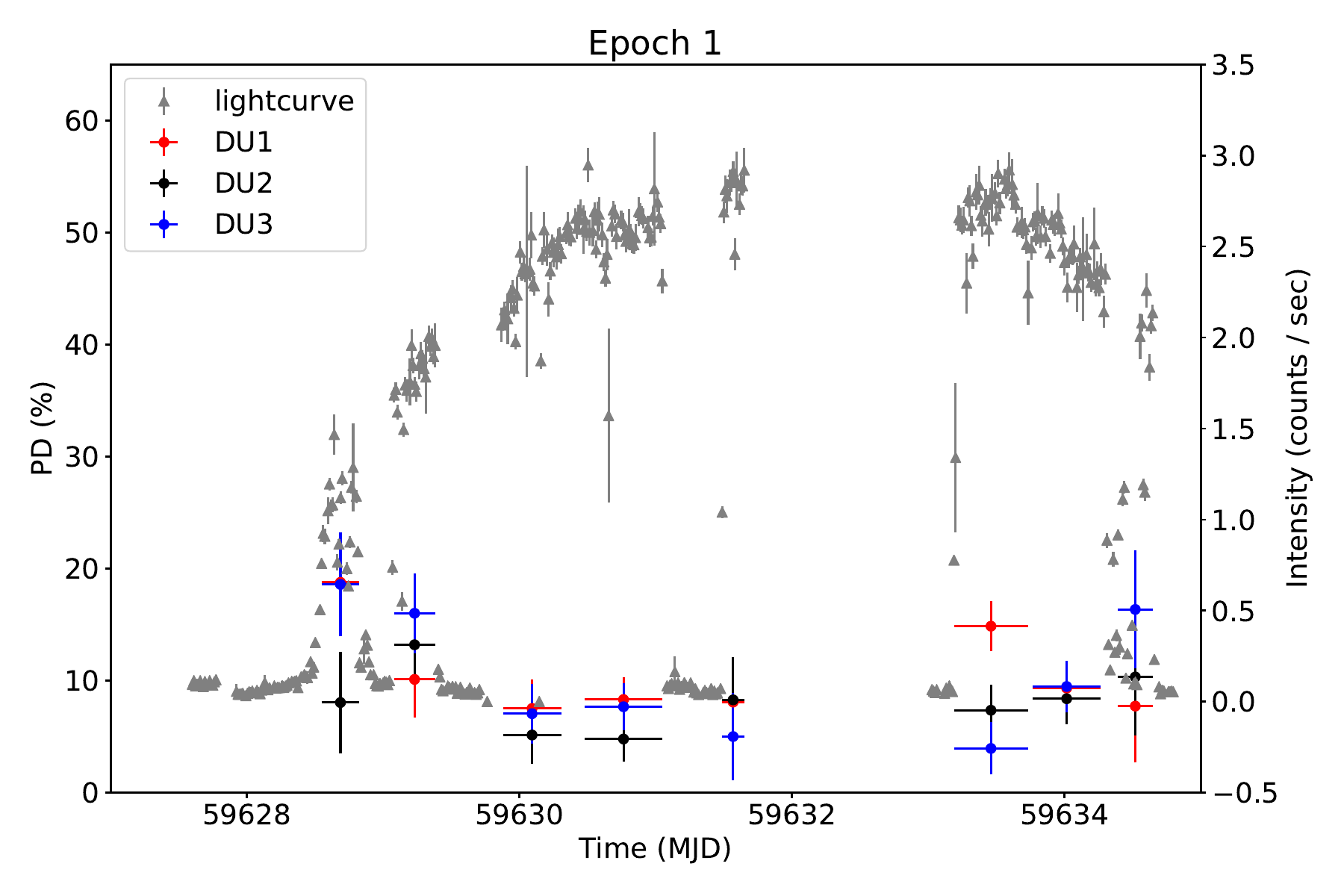}
\includegraphics[width=8.5cm,height=6cm,angle=0]{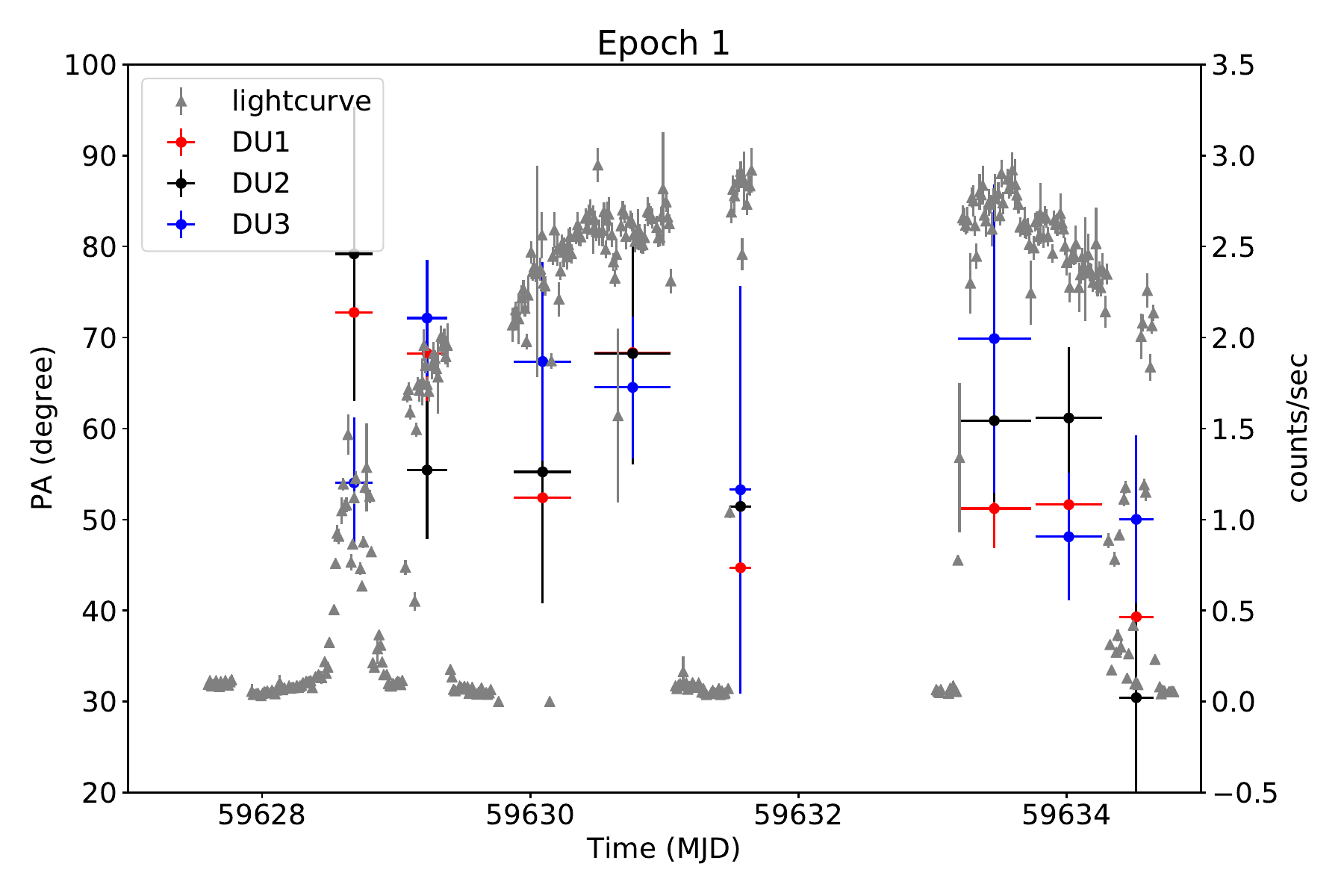}
\includegraphics[width=8.5cm,height=6cm,angle=0]{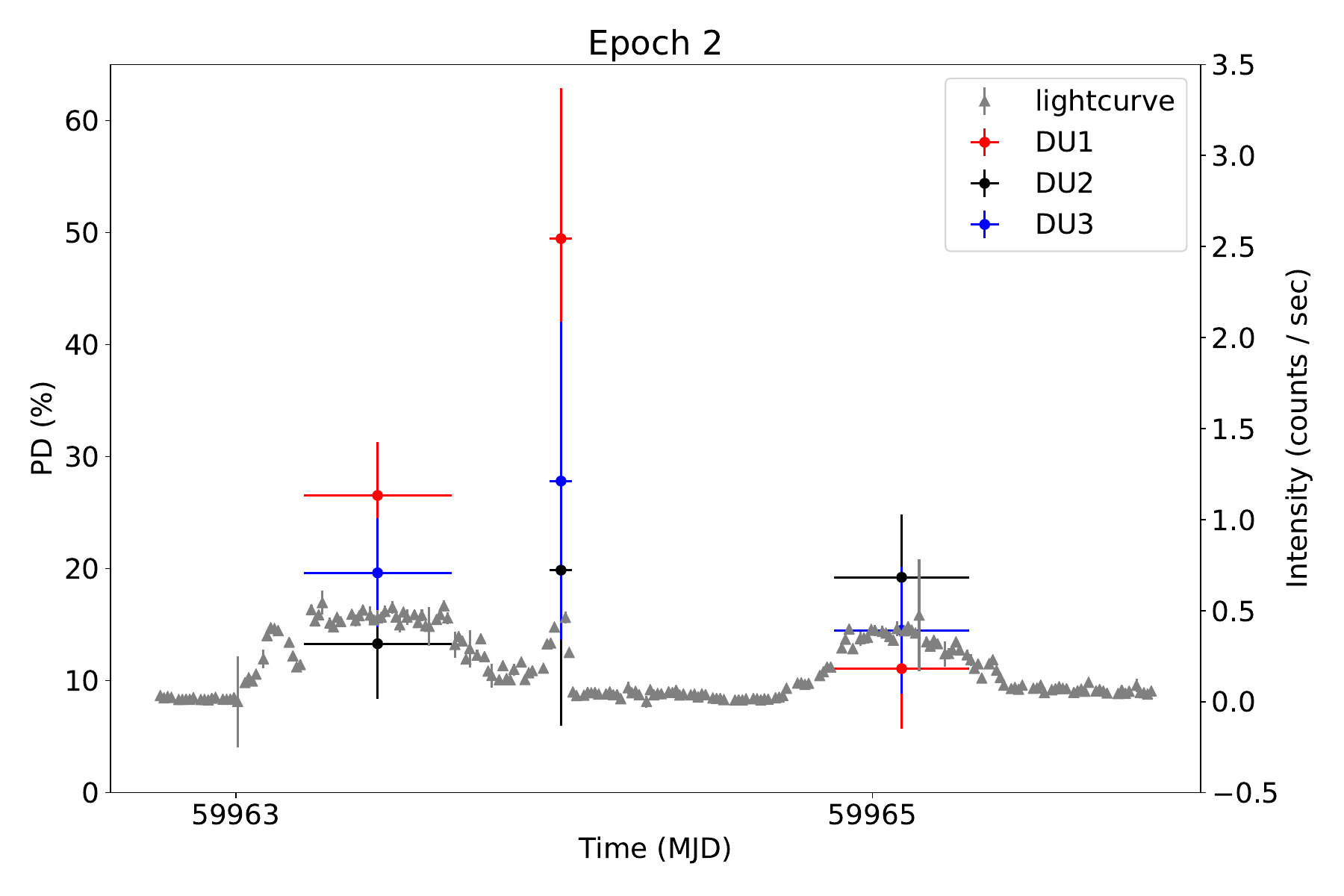}
\includegraphics[width=8.5cm,height=6cm,angle=0]{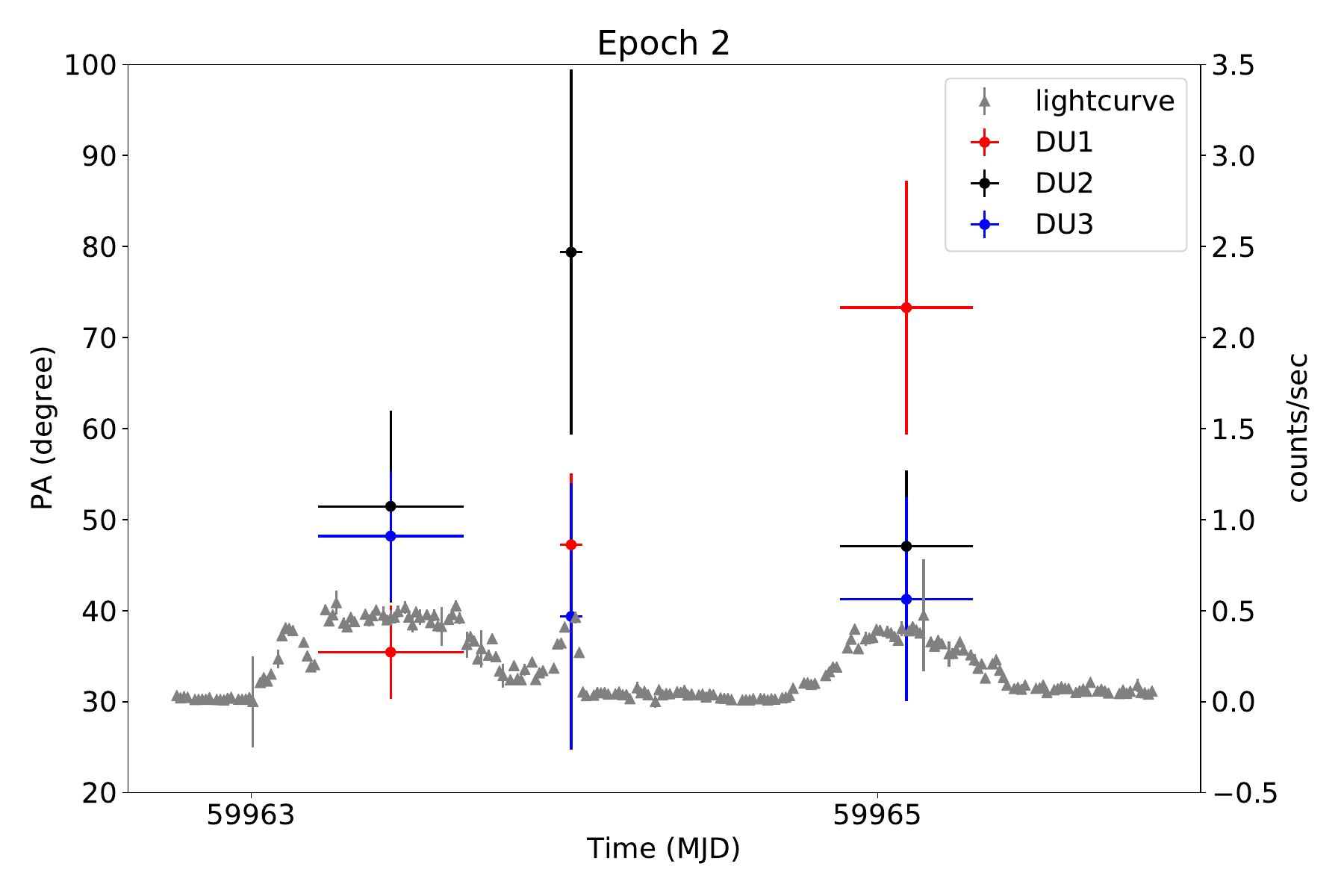}
\includegraphics[width=8.5cm,height=6cm,angle=0]{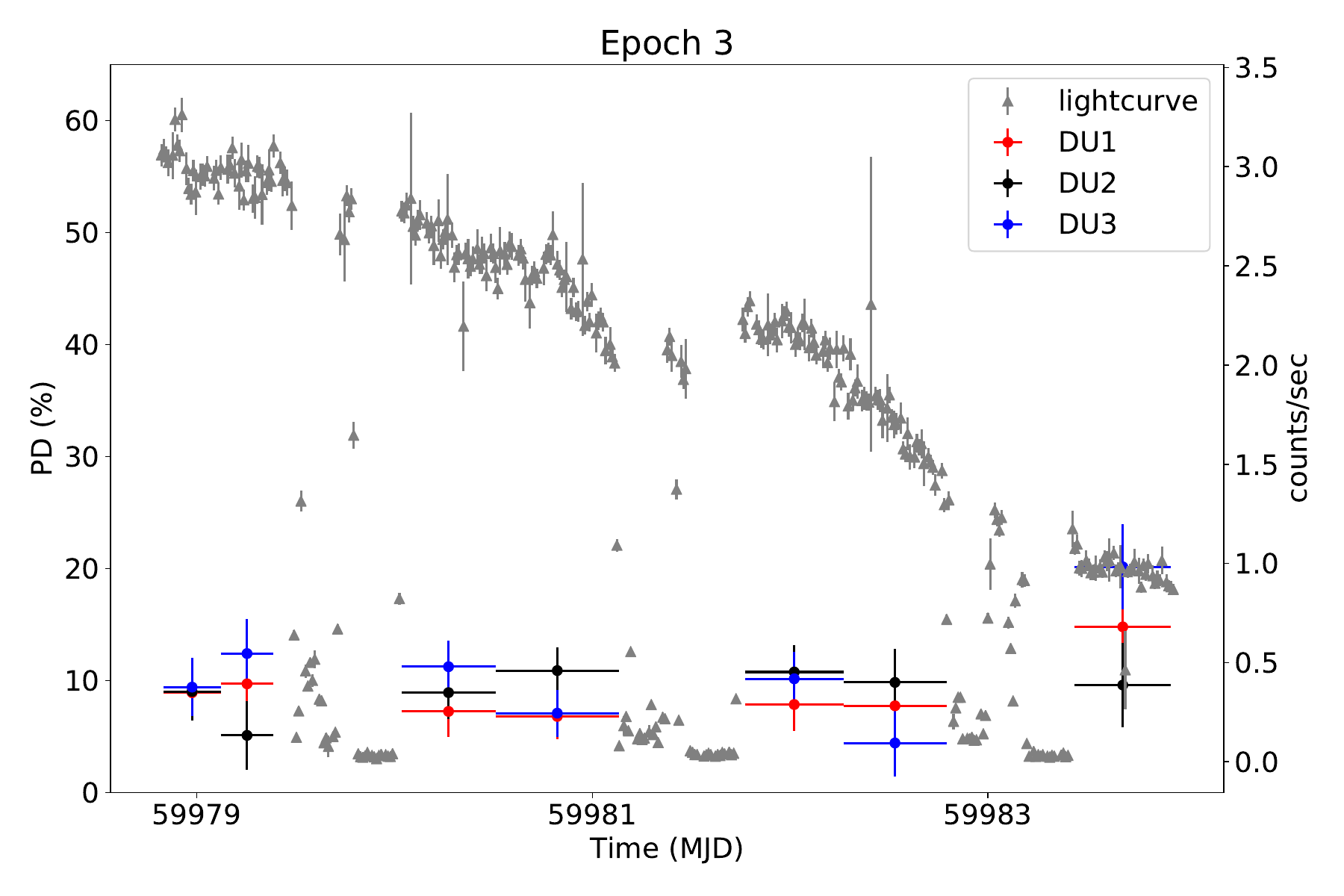}
\includegraphics[width=8.5cm,height=6cm,angle=0]{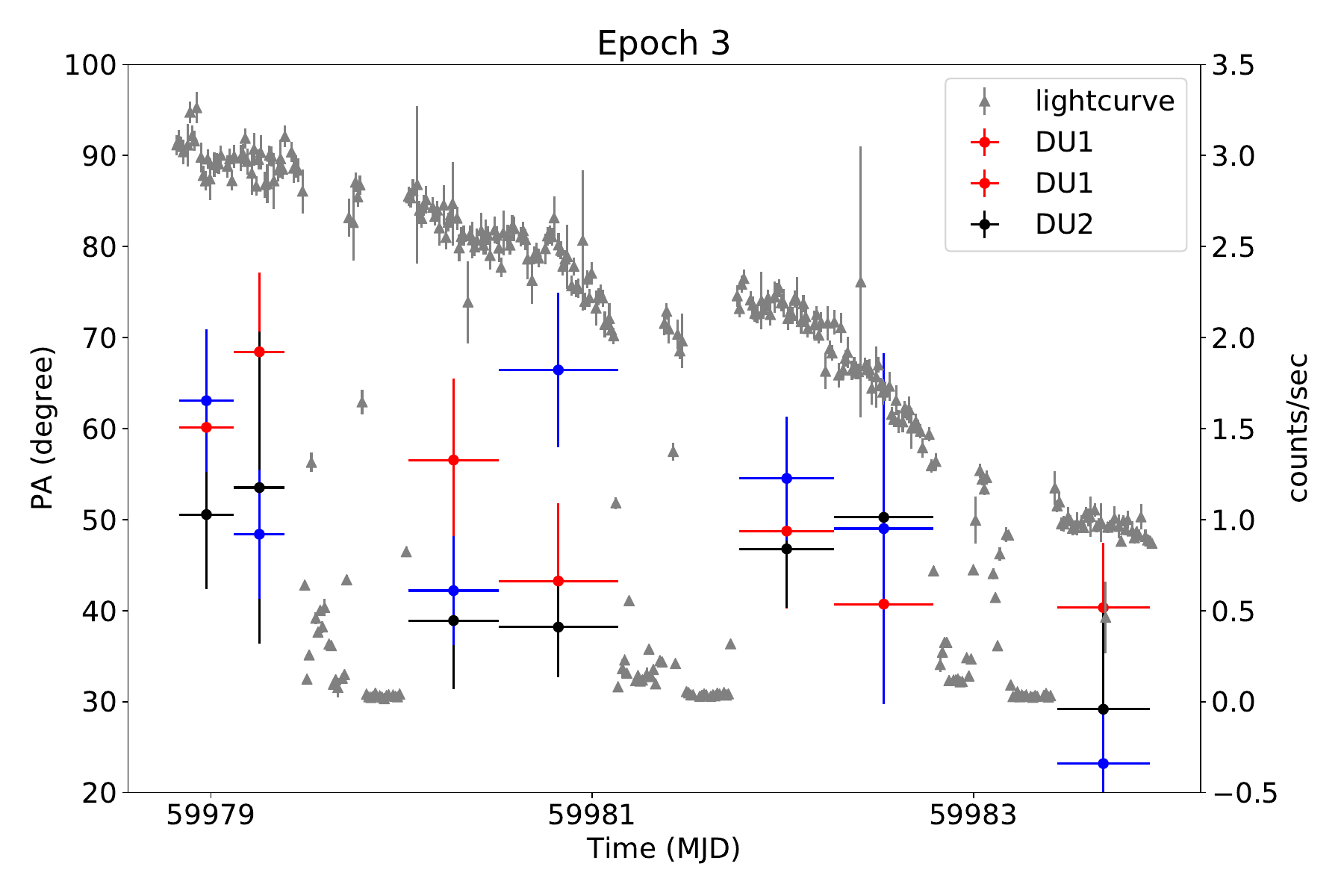}
\caption{PD (left), PA (right), and \textit{IXPE} photon counts (grey points) of \src\ as a function of time for the three epochs. The red, black, and blue symbols represent the data from DU1, DU2, and DU3, respectively.}
\label{pa_pd_lc_seg}
\end{figure*}
%%%%%%%%%%%%%%%%%%%%
%%%%%%%%%%%%%%%%%%%%%%%%%%%%%%%%%%%%%%%%%%%%%%
\begin{figure*}Soft X-ray and FUV observations of Nova Her 2021 (V1674 Her) with
AstroSat
\includegraphics[height=15cm,width=18cm,angle=0]{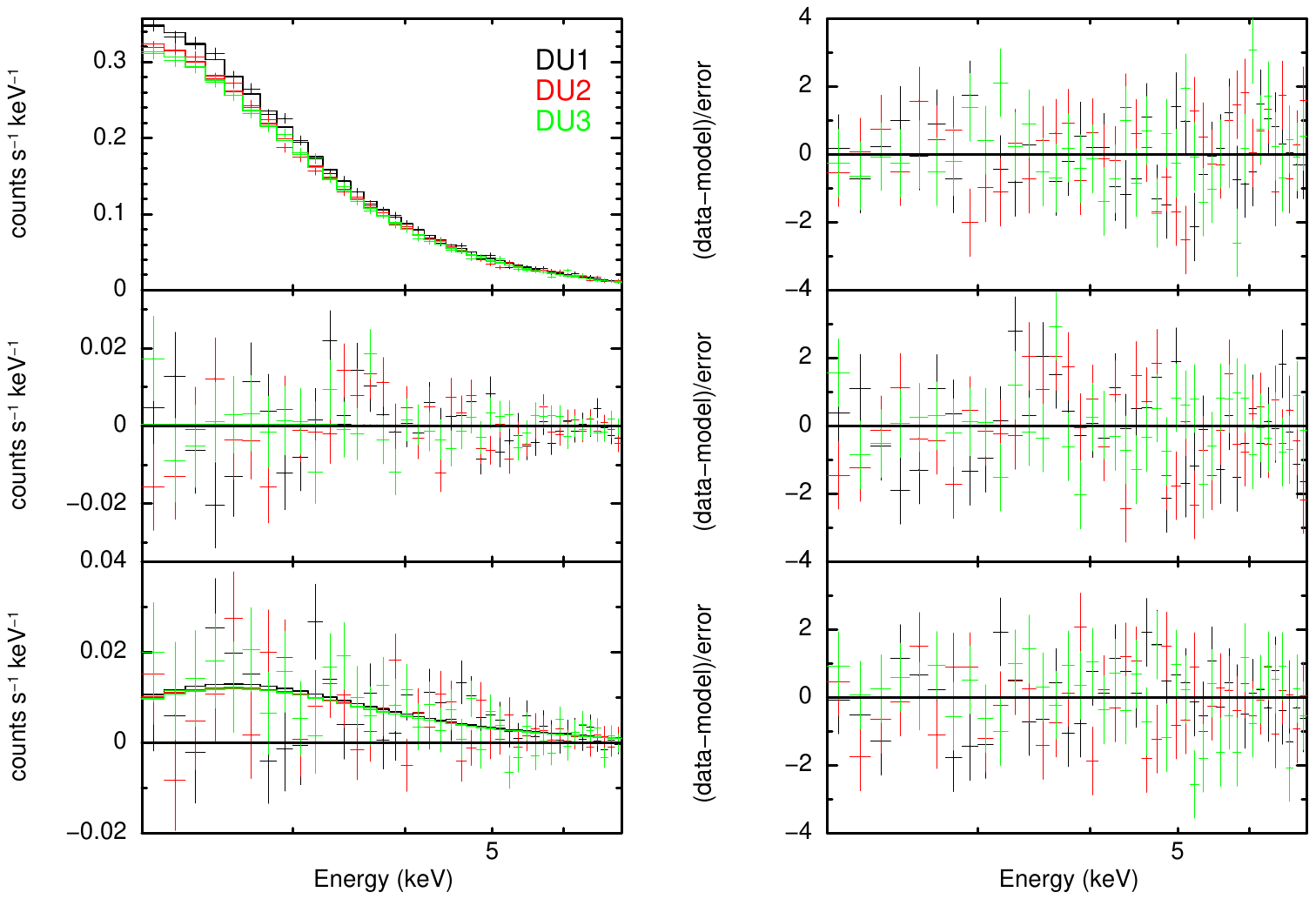}
\caption{The simultaneous {\it IXPE} $I$ (top left panel), $Q$ (middle left panel), and $U$ spectra (bottom left panel) of \src\ fitted with the model  {\sc{constant*polconst*nthcomp}} for Epoch 2. The right panels show the respective residuals of the best-fitting model to the data. The black, red, and green symbols represent the data from DU1, DU2, and DU3, respectively.}
\label{epoch2_sepctra}
\end{figure*}
%%%%%%%%%%%%%%%%%%%%
\section{Results}
\label{Sec_res}

\subsection{Polarization measurements with \textit{IXPE}}
\label{sec_pcube}
Using a model-independent analysis ({\sc pcube} algorithm), we estimate significant polarization in the source at all three Epochs. We find that for Epoch 1; the PD is $7.5 \pm 0.4\%$ at a PA of $59^{\circ} \pm 1.5$ in the $2-7$~keV band for all three DUs. At Epoch 2, we find that the PD has increased to $17.3 \pm 2.0\%$ with the PA decreasing to $46.6^{\circ} \pm 3.4$ (in $2-7$~keV for all 3 DUs) and at Epoch 3 the PD is at $8.2 \pm 0.5\%$ at a PA of $49.8^{\circ} \pm 1.9$. The PD and PA, as seen in the first Epoch, are consistent with \citet{doroshenko2022NatAs...6.1433D} within $2\sigma$ confidence level with the minor differences arising possibly due to the update in the calibration or differences in the choice of the time segments for averaging. The PD increases from Epoch 1 to 2 and then decreases from Epoch 2 to 3, with the values in Epoch 1 and 3 being consistent with each other. We depict the contour plot of the time-averaged polarization parameters, PD, PA, and the Stokes parameters, Q/I and U/I, in the $2-7$~keV energy range for all three Epochs in Figure~\ref{4dparam}. We note that the uncertainties of the PD and PA are calculated assuming that the Stokes parameters Q/I and U/I are normally distributed and uncorrelated and that the PD and PA are independent \citep{2015APh....68...45K}. Table~\ref{PA_PD_obs_log} lists the values of time-averaged PA and PD, combining all detector units at each Epoch.\\

Next, we estimate the time-averaged PA and PD in four energy bands, $2-2.74$~keV, $2.74-3.74$~keV, $3.74-5.12$~keV and $5.12-7$~keV for all DUs to study their energy-dependence as shown in Figure~\ref{ene_pol2}. From this Figure, it is apparent that the polarization parameters are independent or weakly dependent on energy in all three Epochs \citep[similar to ][ for Epoch 1]{doroshenko2022NatAs...6.1433D}.  We also calculate the PA and PD parameters for each segment of all Epochs to examine the time dependence of the observed polarization. Figure~\ref{pa_pd_lc_seg} shows the time evolution of the PA and PD for all three Epochs, with the source light curve shown in grey. For all Epochs, both PA and PD are consistent with a constant throughout each Epoch. This figure also shows an increase in the PD in the low-flux state compared to the two high-flux states; although it is difficult to discern this from these plots due to the large error bars. But the time-averaged results, as seen in Table~\ref{PA_PD_obs_log}, show that the differences are significant. The segment-wise values of the PA and PD for all DUs are given in Table~\ref{PA_PD_log3}.

\subsection{Spectro-polarimetric analysis }\label{spectro-pol}

Besides the {\sc{pcube}} approach to study the polarization properties, we undertake a separate approach as well where we fit the \textit{I}, \textit{Q} and \textit{U} spectra of all \textit{IXPE} detectors in the $2-7$~keV energy band simultaneously in {\sc{xspec}}. Essentially, the idea is to model the \textit{I} spectrum  to determine the radiative mechanism  in conjunction with a polarization {\sc{xspec}} model for the \textit{Q} and \textit{U} spectra to trace the energy dependence of the polarization parameters (PD and PA). The source spectrum in the $2-7$~keV energy band can be well-described by a single power-law with a cutoff at $\sim$20~keV to account for the accretion column emission \citep{kosec2022ApJ...936..185K}. However, \cite{doroshenko2022NatAs...6.1433D} have modeled the radiative mechanism with {\sc{nthcomp}} and found that the polarization is consistent with a constant (on modeling it with {\sc {polconst}}).

In this work, we first fit the time-averaged \textit{I}, \textit{Q} and \textit{U} spectra for all three Epochs using the model {\sc{constant*polconst*powerlaw}}. Since we find that the source polarization is energy-independent (Figure~\ref{ene_pol2}), we model the Stokes spectra using the multiplicative model {\sc{polconst}}. The multiplicative factor {\sc{constant}} takes into account possible energy-independent uncertainties in the cross-calibration of the three {\it{IXPE}} detectors. The spectral fitting gives a $\chi^2$/dof of 713/552, 408/363, and 1154/552 for Epochs 1, 2, and 3, respectively. We keep all the parameters free (i.e. A: the polarization fraction, $psi$: the polarization angle,  $\Gamma$: powerlaw index, and $norm$: powerlaw {\rm Norm}) during the fitting. The high reduced $\chi^2$ of the \textit{I}  spectra suggests that a simple power-law model cannot describe the spectra adequately, and a different model is needed to explain the emission from the accretion column.

Following \cite{doroshenko2022NatAs...6.1433D}, we replace the {\sc powerlaw} component with the component {\sc nthcomp} in {\sc Xspec}, assuming a disk-blackbody seed-photon source; this component computes the Comptonisation of seed photons of temperature $kT_{bb}$ by non-relativistic electrons in a corona of temperature $kT_e$. 
This model yields better spectral fitting with $\chi^2$/dof for the three Epochs of 606/550, 382/361, and 699/550. All the parameters are kept free to vary during the fitting but are linked across the $I$, $Q$, and $U$ spectra. For Epoch 1, we estimate the best-fit values of PD = $8.5\pm0.5 \%$, PA = $59.3^\circ \pm 1.7$, $\Gamma$ = $1.29\pm0.04$, $kT_e$ = $5.75\pm1.55$~keV, $kT_{bb}$ =  $0.58\pm0.03$~keV and $norm$ = $0.111\pm0.002$. All the values except that of $kT_{bb}$ are consistent with the values found by \citet[][see their Supplementary Table~1]{doroshenko2022NatAs...6.1433D}. The best-fit PA and PD values for all Epochs are given in Table~\ref{PA_PD_obs_log}. These values agree with what we get from the {\sc{pcube}} method within $1 \sigma$ errors, thereby confirming that there is indeed a change in the polarization parameters when the source was in the low-flux state. The best fit spectral slope of the spectrum, $\Gamma$, changes from $1.29\pm0.04$ in Epoch 1 to $1.14\pm0.07$ in Epoch 2 and to $1.22\pm0.01$ in Epoch 3. The changes are marginally significant, and the values are consistent within $3\sigma$. Owing to the limited energy range of the data, $2-7$~keV, $kT_e$, and $kT_{in}$ have large error bars and thereby it is difficult to assess whether there is any change. Particularly, $kT_e$ and $kT_{in}$ are respectively measured to be $10.2\pm 14.5$~keV and $<0.63$~keV for Epoch 2 and $7.43\pm 0.25$ and $<0.31$ for Epoch 3. It is important to point out here that the underlying mechanism responsible for the emission of the accretion column in \src\ is likely Comptonization by the bulk motion of electrons which is not the mechanism used by {\sc{nthcomp}}. But as \cite{doroshenko2022NatAs...6.1433D} mentioned, the $\textit{I}$ spectrum in the modest {\it{IXPE}} energy range of $2-7$~keV can be modeled with a simple {\sc{nthcomp}} component only to track the change of the spectral shape between Epochs. In any case, the PA and PD estimates are not affected by the choice of the radiative model.

Figure~\ref{epoch2_sepctra} shows the jointly fitted \textit{I} (top left panel), \textit{Q} (middle left panel), and \textit{U} (bottom left panel) spectra for all three DUs along with their residuals in the adjacent right panels for Epoch 2. 
We also fit the \textit{I} \textit{Q} and \textit{U} spectra of each segment in all Epochs with the model combination {\sc{constant*polconst*powerlaw}} to test whether there was any evolution of PA/PD or the spectral parameters. The reduced $\chi^2$ ranges from 0.98 to 1.14 for Epoch 1, 0.89 to 1.17 for Epoch 2, and 0.84 to 1.17 for Epoch 3. Within $3 \sigma$ limit, the values of PA and PD are consistent with being constant to $\sim 8 \%$ and $\sim 60^\circ$  for Epoch 1, $\sim 18 \%$ and $\sim 45^\circ$  for Epoch 2 and $\sim 10 \%$ and $\sim 50^\circ$  for Epoch 3. This resembles what we found with the {\sc{pcube}} method in Figure~\ref{pa_pd_lc_seg}. The spectral slope seems to vary at $3 \sigma$ in positive correlation with the count rate for Epoch 1 but for Epoch 2 and 3, it is constant around $\sim0.8$ and $\sim 1.0$. Table~\ref{PA_PD_log3} gives the values of all spectral parameters for all segments. 

\section{Discussion and Conclusions}\label{sec_concl}
We present, for the first time, a significant increase in the PD of the accreting X-ray pulsar \src, from $7-9~\%$ in the high-flux state (Main-on) to $15-19~\%$ in the low-flux state (Short-on) within its super-orbital period. For the first Epoch, when the source was in the Main-on phase, the polarimetric and spectro-polarimetric analysis yields a polarization angle of $57-61 ^\circ$. These values are consistent with the  values reported by \citet{doroshenko2022NatAs...6.1433D} for the same observation. For the second Epoch, i.e. the Short-on state,  the polarization angle is $39-50^\circ$. In the third Epoch, when the source is again in the Main-on state, the polarization angle is $47-52 ^\circ $. Within 3$-\sigma$, the polarization angle for all three Epochs is the same. We further find that there is no significant energy dependence of the polarization fraction and polarization angle in any of our Epochs (upper panel of Figure~\ref{ene_pol2}). These results are consistent with \citet{doroshenko2022NatAs...6.1433D} but in contrast with predictions of \citet{ca21}.\\

During the Main-on state, the X-ray emission comes from the neutron star and the tilted accretion disk, whereas in the Short-on state, the disk blocks the neutron star partially and most of the emission originates from the inner regions of the disk \citep{pe91,scott2000ApJ...539..392S,2002MNRAS.334..847L}. Using data from two super-orbital phases, our results show, for the first time, that the PD increases during the low-flux phase of the super-orbital period, i.e., when the central source is more obstructed by the disk warp. The possible reasons for this increase in the PD could be (1) the obscuring structure adds to the PD;
(2) the obscuring structure blocks more the unpolarized than the polarized emission; and (3) the obscuring structure obstructs preferentially the emission from one of the magnetic poles of the NS, thus making the emission more single pole-like and hence increasing the net PD. The first option does not appear to be plausible because if the polarization originates from the accretion column, an obscuring warp, far from the NS, should not increase but decrease the PD. The second option is also unlikely because if we assume that the polarized emission is due to a non-thermal power-law component (which is the only dominant component) in the $2-7$~keV band, then the additional obstruction of an unpolarized component (if any) may therefore not change the PD. Thus, option 3 appears to be the most plausible of the three. If this is the case this result can potentially put a constraint on the geometry of both the warp and the accretion column.\\

Considering the spectral model of \citet{be05}, \citet{ca21}  computed the expected polarized emission of \src, and predicted both phase- and energy-dependent polarization of $60-80$ \% in the 1–10~keV band. On the contrary, using both the model-independent and the spectro-polarimetric methods, here we found that the maximum polarization fraction of \src\ is in the range $15-19~\%$; this high PD was observed during the Short-on state. Our phase-resolved analysis results (see left panel of Figure~\ref{pa_pd_lc_seg} and Table~\ref{PA_PD_log3}) further exclude the possibility that, during these \textit{IXPE} observations, the PD of \src\ was larger than this.
In this context, it is worth developing models of the emission of the accretion column in X-ray binary pulsars. This, plus further observations of the X-ray polarimetric properties of \src\ in other phases of its super-orbital period could shed more light on the accretion geometry in X-ray pulsars.\\

\begin{acknowledgments}
The authors thank the anonymous reviewer for constructive comments to improve the manuscript. This research used data products provided by the \textit{IXPE} Team (MSFC, SSDC, INAF, and INFN). It was distributed with additional software tools by the High-Energy Astrophysics Science Archive Research Center (HEASARC) at NASA Goddard Space Flight Center (GSFC).
MM acknowledges support from the research program Athena with project number 184.034.002, which is (partly) financed by the Dutch Research Council (NWO). MM has benefited from discussions during Team Meetings of the International Space Science Institute (Bern), whose support he acknowledges.
\end{acknowledgments}

%\section{Appendix information}
%\section{Table}
\appendix
\counterwithin{table}{section}
\counterwithin{figure}{section}
\section{Tables and Figures}
\begin{figure*}[h!]
\centering\includegraphics[scale=0.4,angle=0]{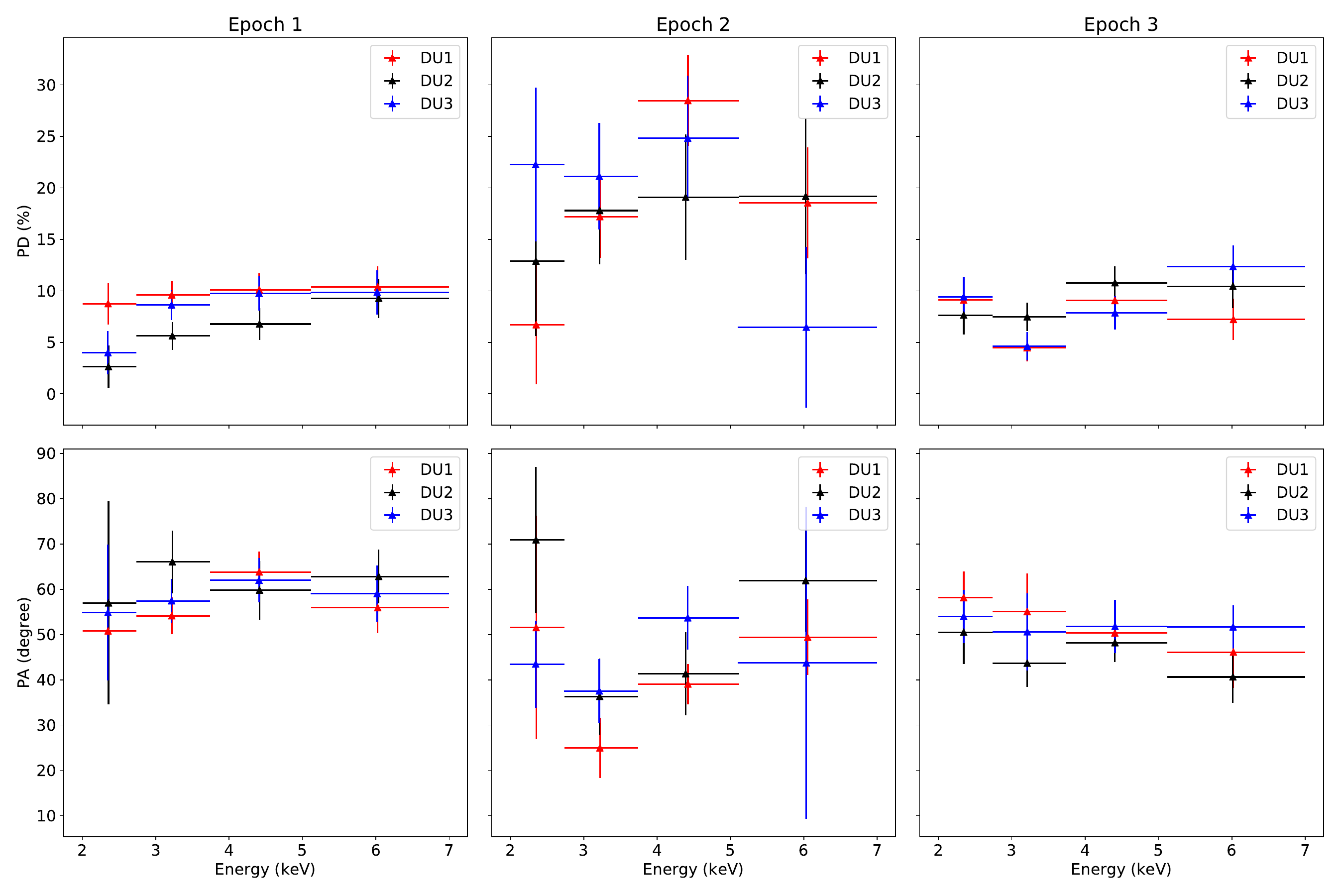}
\caption{The PD and  PA of \src\ as a function of energy for the three Epochs.}
\label{ene_pol2}
\end{figure*}

In this Appendix, we provide the results of the spectro-polarization analysis obtained using the {\sc{pcube}} and {\sc{xspec}} method for each segment of all Epochs in Table \ref{PA_PD_log3}. We also show the evolution of time-averaged PA and PD with energy for all Epochs in Figure \ref{ene_pol2}.

%%%%%%%%%%%%%%%%%%%%%%%%%%%%%%%%

%%%%%%%%%%%%%%%%%%%%%%%%%%%%%%%%%
\begin{table*}[h!]
%\centering
 \caption{Epoch, Observation ID, Start and end time of observations with
 PA and PD using both {\sc{pcube}} and {\sc{xspec}} model {\sc{constant*polconst*powerlaw}}, power law index ($\Gamma$), {\sc powerlaw} {\rm Norm} and goodness of the fit for Her X-1 .}
 \begin{center}
\scalebox{0.9}{%
\hspace{-2.0cm}
\begin{tabular}{ccccccccccc}
\hline
 Epoch & Obs ID  & Tstart & Tstop & \multicolumn{2}{c}{ {\sc{pcube}}} & \multicolumn{5}{c}{\sc xspec}   \\
 & &   (MJD) & (MJD)  & PD (\%) & PA (degree) & PD(\%) & PA (degree)  & $\Gamma$  & Norm   & $\chi^{2}$ (dof) \\ \hline
1 & 01001899
& $59628.55$   &  $59628.82$  & $14.1 \pm 2.6$ & $66.8 \pm  5.3$ &  $14.9\pm{2.3}$  & $66.6\pm{4.4}$  & $0.42\pm{0.02}$ & $1.5\pm{0.0}$ & $594.5 \;(552)$ \\
&  & $59629.08$   &  $59629.38$  & $12.6 \pm  2.0$ & $65.4 \pm 4.6$ & $12.4\pm{1.8}$  & $66.8\pm{4.2}$  & $0.87\pm{0.01}$ & $4.5\pm{0.1}$ & $541.3 \;(552)$ \\
&  & $59629.88$   &  $59630.31$  & $6.4 \pm 1.5$ & $58.1 \pm 6.8$ &  $5.9\pm{1.3}$  & $57.7\pm{6.6}$  & $0.94\pm{0.01}$ & $8.9\pm{0.1}$ & $587.5 \;(552)$ \\
&  & $59630.48$   &  $59631.05$  & $6.9 \pm 1.2$ & $67.0 \pm   4.9$ &  $7.0\pm{1.0}$  & $60.6\pm{4.3}$  & $0.95\pm{0.01}$ & $14.9\pm{0.1}$ & $627.4 \;(552)$ \\
&  & $59631.49$   &  $59631.65$  & $7.1 \pm 2.2$ & $49.2 \pm 8.8$ &  $7.8\pm{2.0}$  & $55.6\pm{7.2}$  & $0.96\pm{0.01}$ & $4.3\pm{0.1}$ & $570.4 \;(552)$ \\
&  & $59633.19$   &  $59633.73$  & $8.6 \pm 1.3$ & $56.5 \pm 4.4$ &  $8.6\pm{1.2}$  & $58.5\pm{3.9}$  & $1.09\pm{0.01}$ & $13.7\pm{0.1}$ & $633.8 \;(552)$ \\
&  & $59633.77$   &  $59634.26$  & $8.9 \pm 1.3$ & $53.5 \pm 4.2$ &  $9.3\pm{1.2}$  & $57.2\pm{3.6}$  & $1.08\pm{0.01}$ & $13.7\pm{0.1}$ & $596.4 \;(552)$ \\
&  & $59634.39$   &  $59634.65$  & $10.8 \pm  3.0$ & $41.5 \pm 7.9$ &  $12.1\pm{2.7}$  & $42.3\pm{6.4}$  & $0.71\pm{0.02}$ & $1.6\pm{0.0}$ & $585.3 \;(552)$ \\ \hline
2 & 02003801
 & $59963.21$   &  $59963.68$  & $19.3 \pm 2.8$ & $43.1 \pm 4.2$ &  $21.1\pm{2.5}$  & $39.8\pm{3.4}$  & $0.96\pm{0.02}$ & $4.5\pm{0.1}$ & $427.9 \;(363)$ \\
&  & $59963.99$   &  $59964.06$  & $28.7 \pm 8.0$ & $60.0 \pm 8.0$ &  $24.1\pm{7.2}$  & $42.7\pm{6.5}$  & $0.77\pm{0.06}$ & $0.4\pm{0.0}$ & $383.3 \;(363)$ \\
&  & $59964.88$   &  $59965.30$  & $13.5 \pm 3.2$ & $51.7 \pm 6.8$ &  $16.1\pm{2.8}$  & $50.6\pm{5.1}$  & $0.74\pm{0.02}$ & $2.6\pm{0.1}$ & $325.8 \;(363)$ \\ \hline
3 & 02004001 
 & $59978.83$   &  $59979.12$  & $9.0 \pm 1.5$ & $58.0 \pm 4.7$ &  $8.2\pm{1.3}$  & $59.7\pm{4.5}$  & $0.99\pm{0.01}$ & $10.4\pm{0.1}$ & $575.8 \;(552)$ \\
&  & $59979.12$   &  $59979.39$  & $ 8.6 \pm 1.7 $& $ 56.8 \pm 5.8$ &  $9.1\pm{1.5}$  & $64.5\pm{4.8}$  & $0.99\pm{0.01}$ & $7.3\pm{0.1}$ & $525.1 \;(552)$ \\
&  & $59980.04$   &  $59980.51$  & $ 8.8 \pm 1.3 $ & $45.1 \pm 4.4$ &  $6.6\pm{1.2}$  & $57.8\pm{5.1}$  & $0.98\pm{0.01}$ & $12.3\pm{0.1}$ & $637.9 \;(552)$ \\
&  & $59980.51$   &  $59981.14$  & $ 7.6 \pm
 1.2$ & $ 47.0 \pm 4.5$ & $7.0\pm{1.0}$  & $49.6\pm{4.3}$  & $0.99\pm{0.01}$ & $15.5\pm{0.1}$ & $650.9 \;(552)$ \\
&  & $59981.77$   &  $59982.27$  & $9.5 \pm  1.4 $ & $50.0 \pm 4.2$ &  $7.4\pm{1.2}$  & $51.6\pm{4.7}$  & $1.01\pm{0.01}$ & $12.1\pm{0.1}$ & $466.9 \;(552)$ \\
&  & $59982.27$   &  $59982.79$  & $7.3 \pm 1.7$ & $46.5 \pm 6.7$ &   $8.3\pm{1.5}$  & $37.0\pm{5.1}$  & $0.98\pm{0.01}$ & $7.6\pm{0.1}$ & $518.6 \;(552)$ \\
&  & $59983.44$   &  $59983.92$  & $14.2 \pm   2.2$ & $30.4 \pm 4.4$ &  $17.1\pm{1.9}$  & $39.8\pm{3.2}$  & $0.93\pm{0.01}$ & $4.4\pm{0.1}$ & $588.4 \;(552)$ \\

 \hline

\end{tabular}}
\label{PA_PD_log3}
\end{center}
\end{table*}

\bibliography{manuscript}{}
\bibliographystyle{aasjournal}

\end{document}